# Electrical half-wave rectification at ferroelectric domain walls

J. Schaab[1], S. H. Skjærvø[2], S. Krohns[3], X. Dai[1], M. Holtz[4], A. Cano[5], M. Lilienblum[1], Z. Yan[6,7], E. Bourret[7], D. A. Muller[4,8], M. Fiebig[1], S. M. Selbach[2], and D. Meier[1,2]

[1] Department of Materials, ETH Zurich, Vladimir-Prelog-Weg 4, 8093 Zurich, Switzerland

[2] Department of Materials Science and Engineering, Norwegian University of Science and Technology (NTNU), 7491 Trondheim, Norway

[3] Experimental Physics V, University of Augsburg, 86159 Augsburg, Germany

[4] School of Applied and Engineering Physics, Cornell University, Ithaca, New York 14853, USA

[5] CNRS, Université de Bordeaux, ICMCB, UPR 9048, 33600 Pessac, France

[6] Department of Physics, ETH Zurich, Otto-Stern-Weg 1, 8093 Zurich, Switzerland

[7] Materials Science Division, Lawrence Berkeley National Laboratory, CA, USA

[8] Kavli Institute at Cornell for Nanoscale Science, Ithaca, New York 14853, USA

email: dennis.meier@ntnu.no

**Ferroelectric domain walls represent multifunctional 2D-elements with great potential for novel device paradigms at the nanoscale. Improper ferroelectrics display particularly promising types of domain walls, which, due to their unique robustness, are the ideal template for imposing specific electronic behavior. Chemical doping, for instance, induces p- or n-type characteristics and electric fields reversibly switch between resistive and conductive domain-wall states. Here, we demonstrate diode-like conversion of alternating-current (AC) into direct-current (DC) output based on neutral 180° domain walls in improper ferroelectric $ErMnO_3$. By combining scanning probe and dielectric spectroscopy, we show that the rectification occurs for frequencies at which the domain walls are fixed to their equilibrium position. The practical frequency regime and magnitude of the output is controlled by the bulk conductivity. Using density functional theory we attribute the transport behavior at the neutral walls to an accumulation of oxygen defects. Our study reveals domain walls acting as 2D half-wave rectifiers, extending domain-wall-based nanoelectronic applications into the realm of AC technology.**

The observation of electrically conducting domain walls in otherwise insulating ferroelectrics[1] has fueled the idea of designing domain-wall-based devices for next–generation nanoelectronics[2,3,4]. Ferroelectric domain walls represent conducting channels which are spatially mobile and which can be created, positioned, and deleted on demand. In recent years, the diversity of ferroic materials with functional domain walls has expanded significantly[5,6,7] and diverse electronic domain-wall properties have been reported, ranging from strongly insulating to highly conducting behavior[8,9,10,11,12]. Furthermore, the electronic domain-wall state can be manipulated by chemical doping[8,13,14,15] and controlled reversibly by external stimuli[16,17,18,19]. This possibility enables applications beyond the use as rewritable conducting channels. In fact, domain walls can emulate the functionality of digital switches and possibly of transistors and gates[16]. Up to date, most related investigations focus on the domain-wall behavior under direct currents (DC). Many electronic components, however, rely on alternating currents (AC). It is thus crucial to understand the conductance of domain walls at AC frequencies. Analogous to conventional 3D systems, domain walls can be expected to exhibit substantially different transport phenomena under AC and DC currents, which is reflected by recent studies on $Pb(Zr_{0.2}Ti_{0.8})O_3$ thin films[20] and $R$MnO$_3$ ($R$ = Y, Er, Ho) bulk systems[21]. It was shown, for instance, that domain wall oscillations at THz frequencies lead to anomalous microwave AC conductivity[21]. In addition to such dynamical effects, *intrinsic* AC properties may arise at low frequencies, i.e., at energies for which the walls remain stationary, being fixed to their equilibrium position.

In this article, we study the AC conductance at stationary neutral domain walls in improper ferroelectric ErMnO$_3$, covering frequencies in the kilo- to megahertz range. In this low-frequency regime, the neutral domain walls are clamped to their equilibrium position, promoting diode-like electronic properties. Similar to semiconducting diodes, electrically contacted walls allow for converting alternating current (AC) to direct current (DC), representing 2D half-wave rectifiers. By performing scanning probe microscopy and macroscopic dielectric spectroscopy, we quantify the characteristic frequencies at which the rectifying domain-wall properties arise and reveal their relation to the conductivity of the surrounding domains. Based on scanning transmission electron microscopy (STEM) and density functional theory (DFT) calculations we conclude that oxygen interstitials tend to accumulate at the neutral walls. The oxygen interstitials are charge-compensated by electron holes and locally enhance the conductivity. Because of the enhanced conductivity, diode-like conversion occurs up to higher frequencies than for the bulk, providing a microscopic explanation for the unusual AC transport behavior at the neutral domain walls.

The multiferroic hexagonal manganite $ErMnO_3$ displays improper ferroelectricity below $T_C \approx$ 1470 K[22]. Driven by a trimerizing lattice distortion, the spontaneous polarization arises as a secondary effect[23]. The system develops all fundamental types of ferroelectric 180° domain walls[4], including neutral (side-by-side) as well as negatively (tail-to-tail) and positively charged (head-to-head) domain walls[9]. While charged walls were studied intensively and their basic transport properties are well understood[9,10,16,24], much less is known about the neutral domain walls, and they were reported to exhibit suppressed[9,25], bulk-like[26], or enhanced DC conductance[27]. As a possible origin for the diverse conductance properties, a correlation to the oxidation state of the host materials was proposed[27], yet without clarifying its microscopic origin.

In order to understand the complex transport phenomena at neutral domain walls and study their AC behavior in the so far unexplored sub-microwave range, we apply a modification of conductive atomic force microscopy (cAFM). Figure 1a illustrates the applied AC-cAFM experiment. The setup allows for measuring the DC response to an AC voltage applied to the sample (see Methods for details). A representative piezoresponse force microscopy (PFM) image[28] and simultaneously recorded AC-cAFM scan gained on a floating-zone-grown $ErMnO_3$ sample with out-of-plane polarization, i.e., (001)-orientation, are presented in Fig. 1b,c, respectively. The images are taken with a voltage amplitude of 3 V at frequency $\nu$ = 0.3 MHz. They show the distribution of the ferroelectric domains (Fig. 1d displays the surface topography). A comparison of Figs. 1b and 1c reveals that $-P$ domains exhibit a higher current than $+P$ domains in AC-cAFM. The result is in agreement with previous DC measurements and can be explained based on the formation of a Schottky-like barrier between the metallic probe tip and the semiconducting sample[25,26]. We observe a higher barrier at the surface of +$P$ domains than −$P$ domains, giving rise to domain-specific antisymmetric $I(V)$-curves as sketched in Fig. 1e. As a consequence, the time-averaged output signal is different for the two domain states (DC response, Fig. 1f), leading to the conductance contrast in Fig. 1c. Most importantly, the frequency of the applied AC voltage can be tuned. This tunability allows for separating different contributions to the conduction – analogous to conventional dielectric spectroscopy[29] – with a lateral resolution of about 20 nm[30].

A frequency-dependent AC-cAFM image series is shown in Fig. 2a-c. We find that the domain-related conductance contrast decreases with increasing frequency. Most interestingly, a distinct signal emerges at the neutral domain walls as the frequency exceeds ≈ 0.5 MHz (Fig. 2b,c). Figures 2a-c thus reveal the existence of two different frequency regimes: (i) A low-frequency regime, where domain contrast dominates and a second one (ii) where only the domain walls exhibit a detectable AC-cAFM signal.

We first focus on the evolution of the domain contrast as function of frequency. A detailed analysis of the AC-cAFM signal in the low-frequency regime (i) is presented in Fig. 2d, evaluated for the cross-section marked in Fig. 2a. Starting from low frequencies, the current, $I_{+P/-P}$, for both +$P$ and −$P$ domains first increases, reaching a maximum at 0.16 MHz. This is the frequency at which the strongest domain contrast is observed. Towards higher frequencies, the domain contrast $\Delta I = I_{-P} - I_{+P}$ quickly decreases, falling below the detection limit (≈10 fA) at $\nu \approx$ 2.1 MHz. This frequency dependence is in good agreement with our macroscopic measurements of the dielectric permittivity $\varepsilon'(\nu)$ in the frequency range from 40 Hz to 110 MHz. The bulk dielectric properties, i.e., the integrated signal from both domains +$P$ and −$P$, reveal a step-like decrease in $\varepsilon'(\nu)$ at about the same frequency as the +$P$/-$P$ domain contrast in the AC-cAFM response vanishes (see inset to Fig. 2d). This behavior is consistent with the aforementioned emergence of a Schottky-like barrier at the contact-sample interface, which acts as a thin insulating layer giving rise to a high capacitance. At sufficiently high frequency[31], this barrier layer is short-circuited. The effect can be fitted by a simple equivalent circuit model (see fits in the inset to Fig. 2d), assuming two *RC*-circuits connected in series as sketched in Fig. 2d[31]. The two circuits represent the residual conductance and capacitance of the barrier layer and the bulk, with $R_{\text{barrier}} \gg R_{\text{bulk}}$ and $C_{\text{barrier}} \gg C_{\text{bulk}}$. The dielectric properties are obtained from this model using the bulk geometry of the plate-like sample. As described in detail by Ruff et al.[29], the static limit of $\varepsilon'(\nu)$ at low frequencies is artificially enhanced up to $\varepsilon' \approx 500$ compared to the intrinsic bulk dielectric constant. The intrinsic bulk dielectric constant is accessible at higher frequencies for which the barrier layer is short-circuited and is in the order of 20. The position of the loss peak tan $\delta$ associated with the step-like decrease in $\varepsilon'(\nu)$ can be assumed to $\nu_{\text{max}} = 1/(2\pi\tau) \approx 1/(2\pi\varepsilon'/\sigma_{\text{DC}}^{\text{bulk}})$, where $\tau$ is the relaxation time of the step-like feature and $\sigma_{\text{DC}}^{\text{bulk}}$ denotes the bulk DC conductivity. In order to adequately define the cut-off frequency $\nu_c$ at which the Schottky-like barrier at the contact-sample interface is short-circuited, we consider the frequency at which tan $\delta$ falls below 25% of the $\nu_{\text{max}}$-value ($\nu_c$ = 2.0 MHz, see inset to Fig. 2d).

To further characterize the relation between the electronic domain properties and the respective cut-off frequency obtained by AC-cAFM and macroscopic dielectric spectroscopy, we use an ErMnO$_3$ sample series with varying bulk conductivity. Platelets with a thickness of about 200 μm are cut from the same ErMnO$_3$ single crystal and are annealed at 1200 °C under oxygen flow for two hours. Subsequently, the samples are cooled with different rates down to 700 °C (0.01, 0.1, 1.0, and 10 °C min$^{-1}$; at 700 °C the heater is switched off), which adjusts the average domain size as investigated in refs. 32 and 33. Aside from the domain structure, the thermal treatment alters the bulk conductivity as shown in Supplementary Fig. S1. In contrast to the domain size, we find that the DC conductivity monotonously decreases with increasing cooling rate, i.e., shorter high-temperature dwell time. This observation demonstrates that the transport behavior is governed by defect chemistry rather than topological phenomena. In this work, however, we investigate relative differences between bulk and domain walls so that we do not pursue the issue of high-temperature defect chemistry further. Future in-depth studies are highly desirable though, as they may reveal additional pathways for tailoring the response at neutral domain walls.

AC-cAFM spectra for two samples with suppressed and enhanced conductivity are presented in Fig. 3a and 3c, respectively. Qualitatively, the spectra show similar features as Fig. 2d, i.e., dominant domain contrast at low frequencies and pronounced domain-wall contrast at higher frequencies. The frequency at which the domain contrast vanishes, however, strongly depends on the bulk conductivity, shifting to higher frequencies as $\sigma_{DC}$ increases. The observed behavior is in agreement with the current, $\varepsilon'(\nu)$, and tan $\delta$ data (Figs. 3b,d), which corroborate that a higher frequency is required to short-circuit the barrier in case of a more conducting sample. The characteristic frequencies extracted from tan $\delta$ and AC-cAFM data are in reasonable agreement, and we find AC-cAFM cut-off frequencies of about 0.11 MHz (Fig. 3a) for the sample with the lower bulk conductivity ($\sigma_{DC}^{bulk}$ = 0.7 · $10^{-6}$ S/cm) and 3.58 MHz (Fig. 3c) for the sample with the higher conductivity ($\sigma_{DC}^{bulk}$ = 19 · $10^{-6}$ S/cm). The results lead to the ($\nu$, $\sigma$)-diagram in Fig. 3e, where we plot all characteristic frequencies derived by AC-cAFM in relation to the bulk conductivity derived from the macroscopic dielectric analysis. Figure 3e reveals a direct correlation between the rectification observed at neutral domain walls and the conductivity of the domains. By adjusting the bulk conductivity, we set onset frequencies in the order of $10^{-3}$ to 1 MHz, above which only the walls contribute to the AC-cAFM signal, acting as 2D half-wave rectifiers.

We next take a closer look at the neutral domain walls and their specific properties that lead to the pronounced AC-cAFM response in frequency regime (ii). In our AC-cAFM experiments, domain-wall signals are detectable towards much higher frequencies than domain contrast, which implies persistent antisymmetric *I*(*V*) characteristics at the walls. As the domain-wall signal occurs at frequencies that are six orders of magnitude lower than the excitation frequency for domain-wall oscillations[21], we can exclude domain-wall dynamics as possible source for the anomalous AC conductivity. Moreover, we observe similar AC effects for samples with in- and out-of-plane polarization (Supplementary Fig. S2), which is strikingly different from the previously reported domain-wall excitations arising at THz frequencies[21]. Figure 4a shows an AC-cAFM spectrum with extended frequency range, recorded on the same sample as in Figs. 1 and 2. While the domain contrast vanishes at $\approx$ 2.1 MHz (see Fig. 2), the domain walls are resolved in AC-cAFM scans up to about 10 MHz with maximum contrast at $\approx$ 7.37 MHz (Fig. 4b). The 5 times higher cut-off frequency for domain-wall rectification (inset to Fig. 4a) implies a 5 times higher loss-peak position $\nu_c = 1/(2\pi \varepsilon'_{wall}/\sigma_{DC}^{wall})$ than for the bulk and, hence, a substantially higher conductivity ($\sigma_{DC}^{wall} > \sigma_{DC}^{bulk}$) and/or lower intrinsic dielectric constant ($\varepsilon'_{wall} < \varepsilon'$) at the wall.

To understand the microscopic origin of the anomalous transport at the neutral domain walls, we analyze the domain-wall structure at the atomic scale using TEM and DFT. The high-angle annular dark field STEM image in Fig. 5a shows a neutral domain wall, visible as a deviation from the displacement patterns of the erbium ions in $-P$ and $+P$ domains as explained, e.g., in ref. 34. Density functional calculations show that the electronic structure and band gap at such walls are very similar to the bulk when considering a defect free sample (Supplementary Fig. S3). Moreover, we do not find any indication of a locally reduced dielectric constant that may explain the observed shift in the loss peak position at neutral domain walls (Supplementary Fig. S4). The STEM data in Fig. 5a,b, however, clearly shows structural changes with the amplitude $Q$ and phase $\varphi$ of the ferroelectric order parameter varying smoothly across the neutral domain walls (see refs. 34 and 35 for details). The associated strain field represents a possible driving force for the formation or accumulation of defects. To explore whether the same oxygen defects that determine the semiconducting bulk properties[36] also play a role for the anomalous transport behavior at the neutral domain walls, we compare their stability at domain walls and in the bulk.

The formation energy of oxygen defects is quantified by considering defects at several structurally stable positions in a 5x2x1 supercell of $YMnO_3$ as shown in Fig. 5c-f ($YMnO_3$ is isostructural and chemically similar to ErMnO$_3$). The supercell in Fig. 5c contains the two possible types of neutral domain walls[37], corresponding to Y1- and Y2-terminated domain walls, respectively. We note that the walls are not atomically sharp; due to the gradual symmetry breaking to $P3c1$ the walls display a finite structural width (Fig. 5d), which is in excellent agreement with the STEM data in Figs. 5a and 5b. Most interestingly, we find a reduced formation energy $E^f$ for interstitial oxygen O$_i$ at the neutral domain walls, whereas oxygen vacancies are predicted to not accumulate at the walls (Supplementary Table T1). Figure 5e reveals a 30 meV lower energy per O$_i$ forming at a wall compared to in the bulk. This energy lowering originates in the discontinuity in the Y corrugation pattern (Fig. 5f), leading to a higher effective Y coordination number for O$_i$ and, hence, higher stability for interstitial oxygen than in the bulk. Differences in $\Delta E^f$ occur between the two walls in Figs. 5c-f because of the differently Y-terminated domains, which create a chemically different O$_i$ environment. We note that the proposed accumulation of O$_i$ alters the electronic structure compared to the idealized defect-free case (Supplementary Fig. S3) as discussed in ref. 36. The latter is expected to also apply to charged domain walls in hexagonal manganites due to their local strain field[34], providing an additional handle for controlling their transport properties beyond the previously discussed electrostatics-driven electronic reconstruction phenomena[4,16]. In summary, DFT reveals that the formation of oxygen interstitials is energetically favored at neutral domain walls, while vacancies show no tendency to accumulate (not shown). Local charge neutrality at neutral domain walls implies that each oxygen interstitial is compensated by two holes[36], which locally enhances the density of hole carriers. As a consequence, the conductivity at neutral domain walls is enhanced ($\sigma_{DC}^{wall} > \sigma_{DC}^{bulk}$), being in agreement with the higher frequency required to shorten the associated Schottky-like barrier. We thus conclude that oxygen interstitials are responsible for the rectifying domain-wall behavior observed in our AC-cAFM scans, corroborating the previous assumption that the transport at neutral domain walls is determined by the oxygen stoichiometry of the host material[27].

The neutral domain walls investigated in this work promote rectifying behavior at frequencies in the kilo- to megahertz range, facilitating AC-to-DC conversion (half-wave rectifiers) at the atomic scale. Going beyond standard metal-semiconductor and semiconductor-semiconductor junctions in layered materials, the involvement of domain walls enables nano-diodes with minimum lateral size, which is defined by the smallest possible contact area ($\approx$ 1 nm for commercially available probes). Our microscopic model suggests that the walls owe their unusual properties to charge compensated oxygen interstitials, which allows controlling key parameters such as processible input frequencies and magnitude of the output via the oxygen content of the host material. In general, an expansion of domain-wall-based nanoelectronics into the realm of AC currents is appealing as it enables, e.g., the design of nanoscopic capacitors, inductors, and transformers. An involvement of AC currents is also more energy efficient compared to DC currents as smaller currents can be used. The latter significantly reduces heat generation, which is crucial for the design of miniaturized domain-wall-based devices and the downscaling electronic circuits to the level of domain walls.

## Methods

**AC conductive atomic force microscopy (AC-cAFM).** Spatially resolved AC-transport measurements were carried out on an NTEGRA *Prima* scanning probe microscope from NT-MDT. For stable topography and current imaging, the samples were fixed with silver paint to metal disks and mounted onto the bottom scanner of the microscope. The configuration of the microscope for electrical measurements is presented in Figure 1a. AC voltages with frequencies ranging from 0.02 to 20 MHz and an amplitude of 3 V are applied to the back-electrode of the sample (AC input in Fig. 1a). DC contributions of the AC current provoked by asymmetric *I*(*V*) characteristics are measured with the current amplifier directly connected after the cantilever with 10 fA precision. Scanning rate was fixed to 2 msec/point, guaranteeing averaging of the signal in time over at least 40 voltage periods.

**Macroscopic dielectric spectroscopy.** Dielectric spectroscopy was performed with an Agilent 4294A operating with frequencies of 40 Hz to 110 MHz at ambient temperatures. For high accuracy, the samples were contacted with brass plates directly at the measurement extension port.

**Scanning transmission electron microscopy.** $ErMnO_3$ platelets were oriented by Laue diffraction such that the specimen would be imaged down the $[1\bar{1}0]$ zone axis of the crystal in STEM. Cross-sectional STEM samples were prepared using an FEI Strata 400 Focused Ion Beam with a final milling step of 2 keV to minimize surface damage. The STEM specimens were imaged on a 100 keV Nion UltraSTEM optimized for EELS imaging (30 mrad convergence angle, 80-130 pA of usable beam current, ≈1 Å spatial resolution).

**Density functional theory.** DFT calculations were performed with the VASP code[38,39] with the PBEsol functional[40] and the spin polarized GGA + U implementation of Dudarev[41] with U = 8 eV applied to the Mn 3d states. The Y(4s,4p,4d,5s), Mn(3s,3p,3d,4s) and O(2s,2p) states were treated as valence electrons in the projector augmented wave method[42] with a plane wave cutoff energy of 550 eV. Brillouin zone integration was performed on a 1x2x2 mesh for the 5x2x1 supercell with 300 atoms. A collinear A-type antiferromagnetic order was imposed on the Mn sublattice with no change across the domain walls. The geometry was optimized until forces on the ions were smaller than 0.02 eV/Å. VESTA[43] was used to visualize the structures. Defect formation energies were calculated for neutral cells according to $E^f = E_{\text{YMnO}_{3+\delta}} - E_{\text{YMnO}_3} - \mu_{\text{O}}$, with a chemical potential of oxygen $\mu_\text{O}$ = -4.5 eV.

**Acknowledgement:** The authors thank N.A. Spaldin. S.V. Kalinin, and E. Soergel for fruitful discussions. D.M. and J.S. acknowledge funding from the SNF (proposal Nr. 200021_149192) and NTNU`s Onsager Fellowship Program (D.M.). Z.Y and E.B. were supported by the U.S. Department of Energy, Office of Science, Basic Energy Sciences, Materials Sciences and Engineering Division under Contract No. DE-AC02-05-CH11231 within the Quantum Materials program #KC2202. Electron microscopy work was supported by the U.S. Department of Energy, Office of Basic Energy Sciences, Division of Materials Sciences and Engineering under award no. DE-SC0002334. This work made use of the electron microscopy facility of the Cornell Center for Materials Research Shared Facilities which are supported through the NSF MRSEC program (DMR-1719875). S.K. acknowledges funding from the DFG via the Transregional Collaborative Research Center TRR 80 (Augsburg/Munich/Stuttgart, Germany), and from the BMBF via ENREKON 03EK3015.

**Author Contributions:** J.S conducted the SPM study under supervision of D.M. and with assistance from X.D. and M.L., supervised by D.M and M.F, respectively. S.H.S. performed the DFT calculations supervised by S.M.S. S.K. recorded the macroscopic dielectric spectroscopy data. M.H. obtained the STEM data supervised by D.A.M. Z.Y. and E.B. grew the ErMnO$_3$ crystals, and M.L. prepared the series of sample with different conduction properties. J.S., S.K., A.C., and D.M. analyzed the experimental data. D.M. initiated and coordinated this project, and wrote the manuscript supported by J.S., S.K., and S.M.S. All authors discussed the results and contributed to their interpretation.

**Figure Captions**

**Figure 1 | Probing half-wave rectification at the nanoscale. a**, Configuration of the AFM setup used to perform AC-cAFMs measurements. An alternating voltage with frequency $v$ and amplitude $U$ = 3 V is applied to the back-electrode of the sample (AC Input). The resulting DC currents are measured through the AFM probe-tip. **b**, Piezoresponse force microscopy (resonant PFM, out-of-plane contrast), revealing the characteristic domain pattern of hexagonal manganites with six-fold meeting points. **c**, AC-cAFM response recorded at contact-resonance with a frequency of 0.3 MHz, coinciding with the domain pattern resolved by resonant PFM in **b**. **d**, The topography signal recorded at the same position as **b,c** shows no significant spatial variations in the surface structure. **e,f,** Illustration showing how AC voltages lead to DC-current responses due to the asymmetric $I(V)$-characteristics of (001)-oriented $ErMnO_3$.

**Figure 2 | AC-to-DC conversion at domains and domain walls. a**, AC-cAFM image of (001)-oriented $ErMnO_3$ recorded at $v$ = 0.2 MHz, $U$ = 3 V. Domains of opposite polarization exhibit a pronounced contrast with $I_{-P} > I_{+P}$. **b**, Same position as in **a** imaged at higher frequency (1.0 MHz). In addition to the domain-related response, a second signal arises at the position of the domain walls. **c**, At $v$ = 1.61 MHz domain-wall contributions dominate the AC-cAFM data. **d**, Evolution of the AC-cAFM response averaged over the domain-wall section marked in **a** as function of frequency. Domain currents first rise with increasing frequency and then peak at 0.16 MHz, vanishing towards higher frequency. Domain-wall contrast becomes visible at about 0.20 MHz. The inset shows the dielectric permittivity $\varepsilon'(v)$ and loss tangent tan $\delta$ obtained by macroscopic dielectric measurements averaging both $\pm P$ domains. Black solid lines are fits describing the data in terms of an equivalent circuit model, i.e., two $RC$-circuits connected in series (sketch in **d**), and grey dashed line highlight the cut-off frequency $v_c$ as defined in the main text.

**Figure 3 | Relation between bulk conductivity and rectifying domain-wall behavior. a**, AC-cAFM spectroscopy data gained on neutral domain walls in ErMnO$_3$ with reduced bulk conductivity $\sigma_{DC}^{bulk}$ = 0.7 · 10$^{-6}$ S/cm) with $U$ = 3 V. **b**, Frequency dependent dielectric permittivity $\varepsilon'(\nu)$ and loss tangent tan $\delta$ recorded by macroscopic dielectric measurements. Fits to the data (black solid lines) are gained using an equivalent circuit model as described in the main text. The grey dashed lines indicate the cut-off frequency $\nu_c$ = 0.1 MHz (see main text for details). **c,d**, Same as **a,b** for ErMnO$_3$ with higher bulk conductance ($\sigma_{DC}^{bulk}$ = 19 · 10$^{-6}$ S/cm). To describe the dielectric response, an additional RC-element is required compared to **d**, taking into account a second barrier-layer as explained in ref. 29. For the present study, however, only the Schottky-barrier layer contribution at higher frequency is of relevance, describing the frequency above which the intrinsic bulk properties become accessible. The respective cut-off frequency is $\nu_c$ = 6 MHz. **e**, Relation between rectification behavior and bulk conductivity. Blue data points represent the frequency at which domain currents extenuate, i.e., Δ$I$ = $I_{-P} - I_{+P}$ falls below the resolution limit (≈ 10 fA). Orange data points mark the frequency at which domain walls become visible in our AC-cAFM data. Two distinctly different regions are observed in the investigated ($\nu,\sigma$)-regime: A low-frequency regime (i) with dominant domain contrast and a second one at higher frequencies (ii) where only domain walls are detected. In the shaded area both domains and domain walls exhibit rectifying properties.

**Figure 4 | Electrical rectification at neutral ferroelectric domain walls. a**, Extended AC-cAFM spectrum of a neutral domain wall recorded on the same ErMnO$_3$ specimen investigated in Figs. 1,2. The domain-wall spectrum is averaged over the region marked **b** and exhibits similar spectral features as the spectra gained within ±$P$ domains. The maximum current and cut-off frequency, however, are shifted towards significantly higher energy at the walls. The inset illustrates the diode-like rectification properties of the domain wall. From **a**, we find that the cut-off frequency at the walls is about 5 times

higher than for the bulk. **b**, AC-cAFM image taken at 7.37 MHz. White dashed lines mark the region considered in **a**. .

**Figure 5 | Accumulation of oxygen interstitials at neutral ferroelectric domain walls. a**, HAADF-STEM image of a neutral ferroelectric domain wall in ErMnO$_3$, viewed along the P6$_3$cm [1$\bar{1}$0] zone axis, with color overlay of the order parameter phase $\phi$. The color overlay on the brighter Er atomic positions reflect the orientation of the ferroelectric polarization as indicated by the white arrows[34]. **b**, Evolution of the order parameter amplitude $Q$ and phase $\phi$ as function of position relative to the center of the neutral domain wall in **a**. The structure gradually changes across the domain wall, which implies the presence of local strain fields. **c**, Illustration of the YMnO$_3$ supercell for DFT calculations (**d-f**), including two neutral domain walls indicated by purple (Y2-terminated) and orange (Y1-terminated) vertical lines. The shaded areas close to the domain walls indicate the extent of structural distortion. The Y1 atoms are colored blue, and the Y2 are colored green to simplify visualization of the trimerization. The orange spheres show the different stable positions of O$_i$ in the cell. **d**, Order parameter amplitude $Q$ and phase $\phi$ across the stoichiometric supercell in **c**. $Q$ is here represented by the MnO$_5$ trigonal bipyramid tilt angle $\alpha_A$ with respect to the *ab*-plane, while the phase $\phi$ refers to the angle of the trimerization distortion[35]. **e**, Formation energy $E^f$ of oxygen interstitials O$_i$ in stable positions[36] across the supercell in **c**. **f**, Bond distances, Y-O$_i$, between nearest-neighbor Y and oxygen interstitials (O$_i$) across the supercell in **c**.

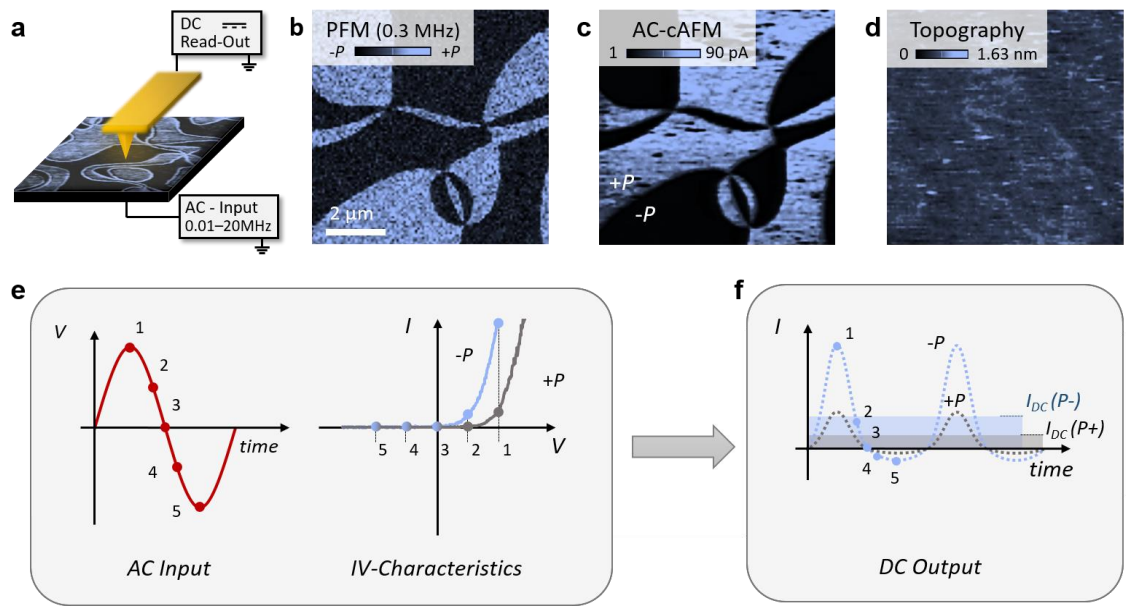

**Figure 1**

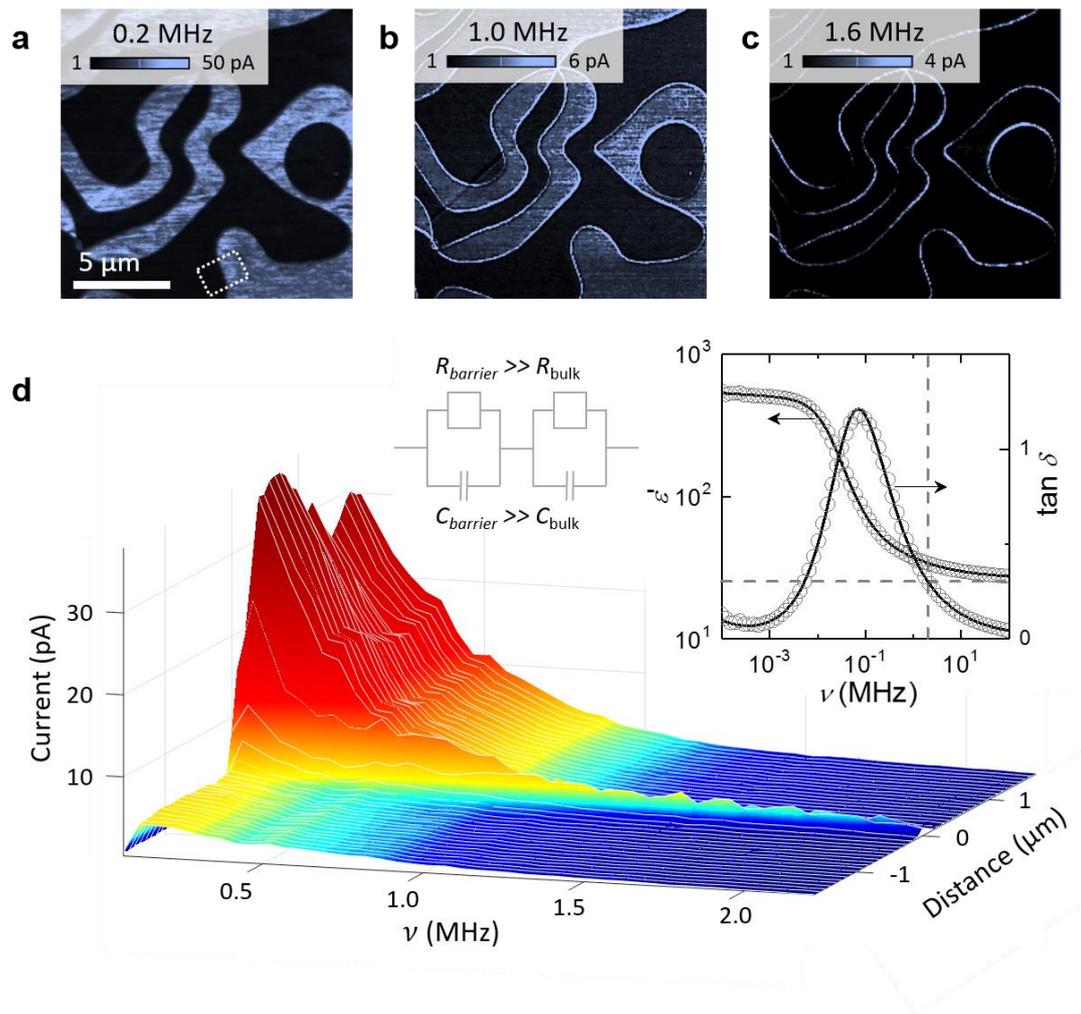

**Figure 2**

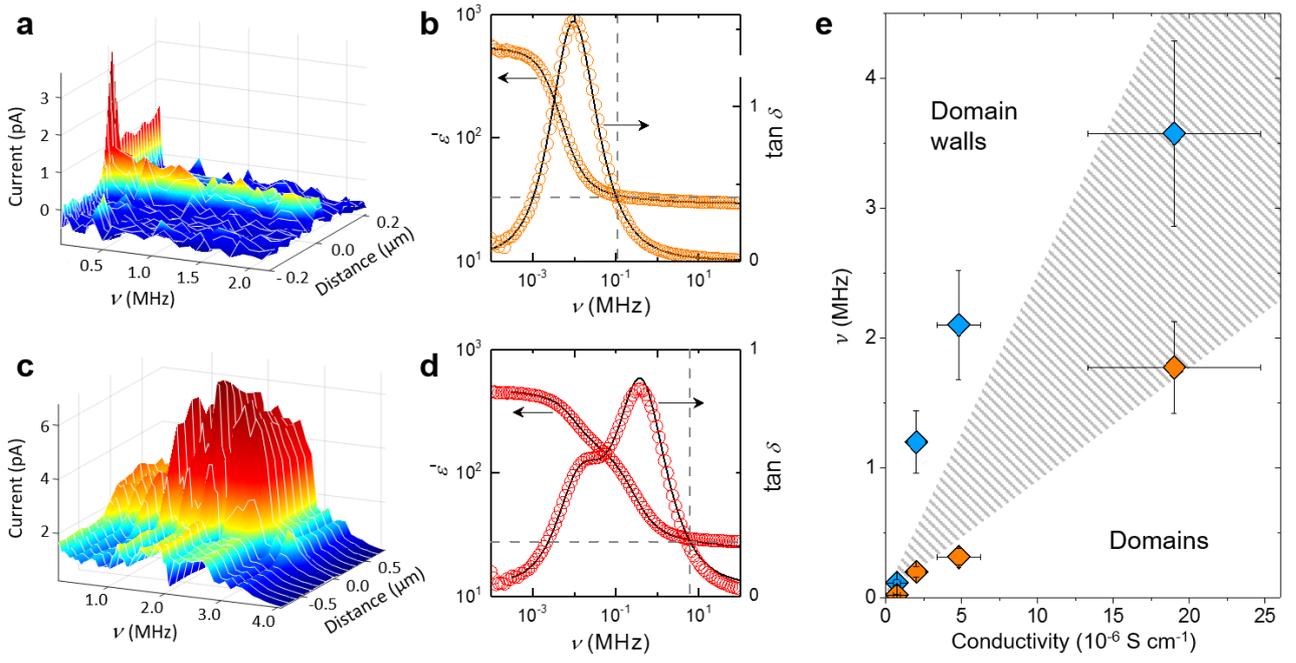

**Figure 3**

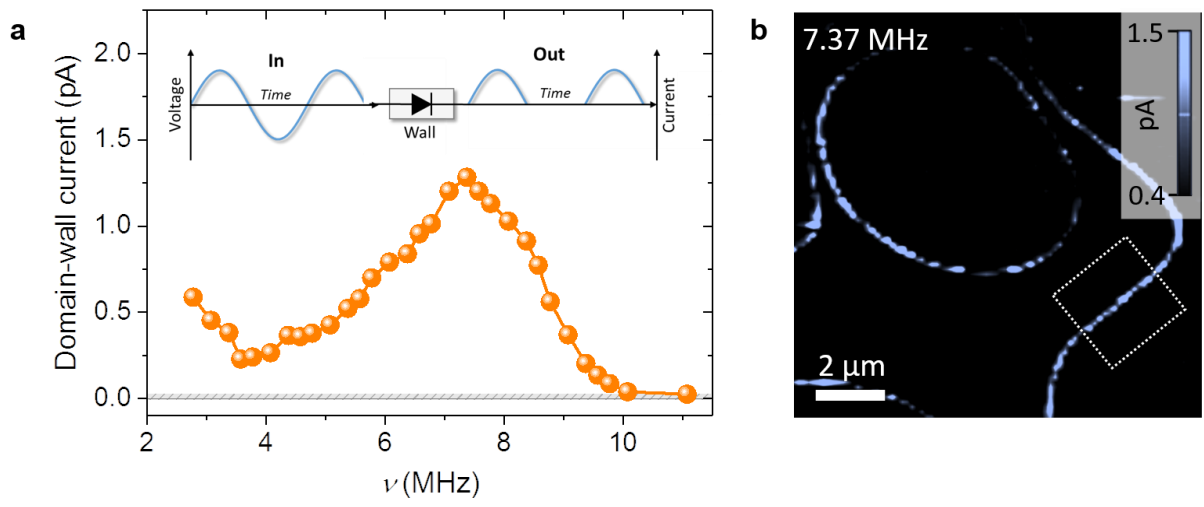

**Figure 4**
23

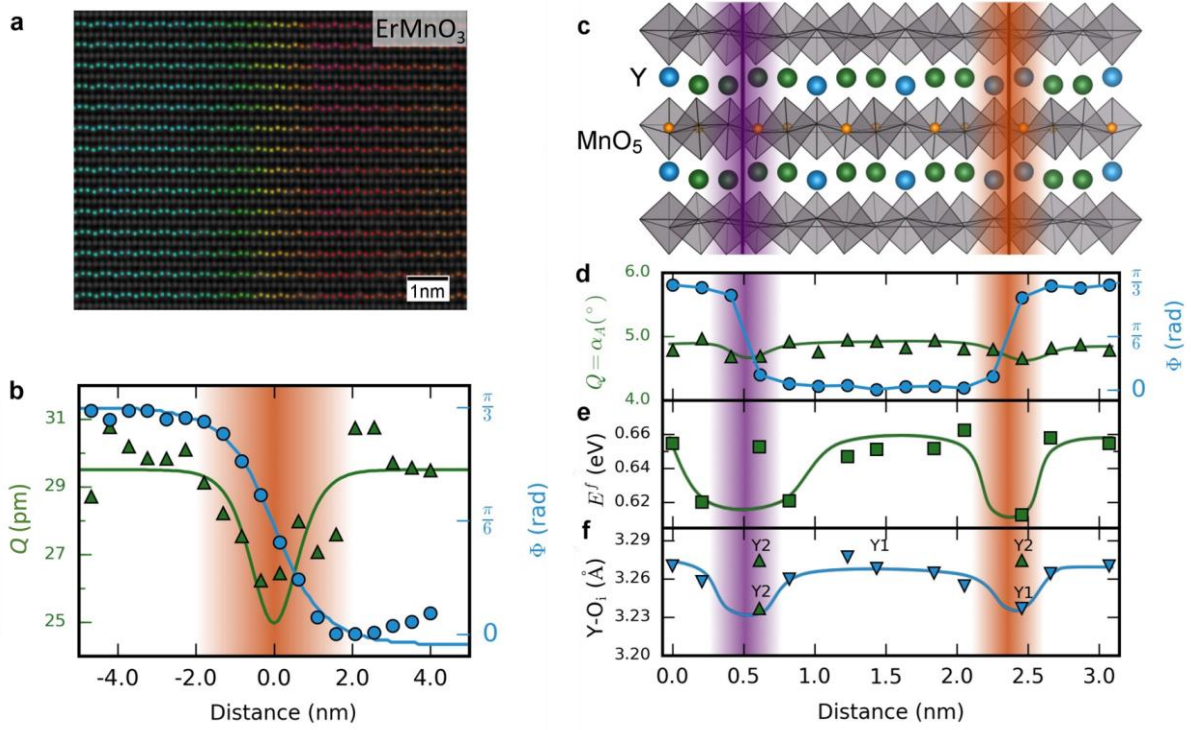

**Figure 5**

# Supplementary Information

## 1. Supplementary Figures

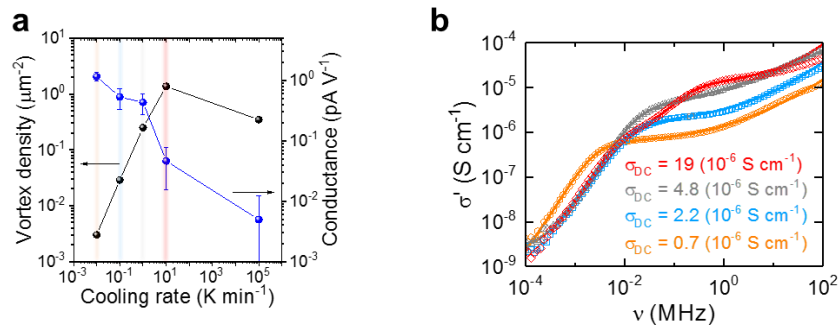

**Figure S1 | ErMnO$_3$ sample series with varying bulk conductivity. a,** Relative bulk conductance (measured by cAFM) and vortex density as function of cooling rate. In contrast to the vortex density (values taken from Ref. 1), the conductance monotonously decreases with increasing cooling rate, reflecting the independence of the two quantities. Marked data points correspond to the samples presented in the ($\nu,\sigma$)-diagram in Fig. 3e in the main text. **b**, Bulk conductance measured by macroscopic dielectric spectroscopy. Solid lines correspond to fits, generated by describing the system by an equivalent circuit model as sketched in the inset to Fig. 2d in the main text.

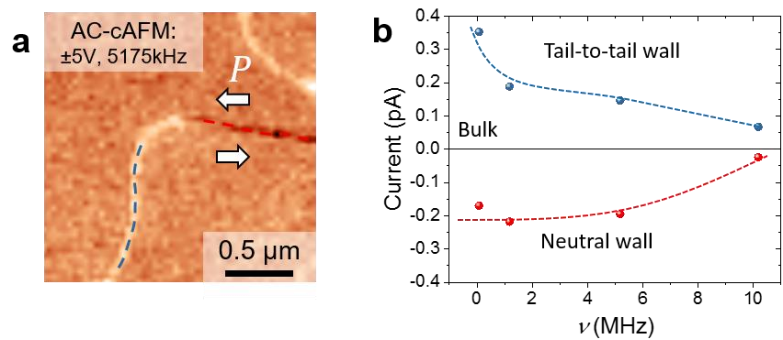

**Figure S2 | Electrical rectification at neutral domain walls in ErMnO$_3$ with in-plane polarization. a**, Both tail-to-tail and neutral domain walls in ErMnO$_3$ with in-plane polarization $P$ exhibit rectifying properties in AC-cAFM scans. The observation of AC-to-DC conversion for in-plane $P$ excludes that the anomalous AC transport is due to tip-induced domain wall movements / displacement current, corroborating its intrinsic nature. **b**, Spectral dependence of the AC-cAFM signal for the two domain walls in **a** for frequencies up to 10 MHz. Blue (red) data points correspond to the maximum (minimum) intensity observed along the domain walls marked in **a**, dashed lines are guides to the eye.

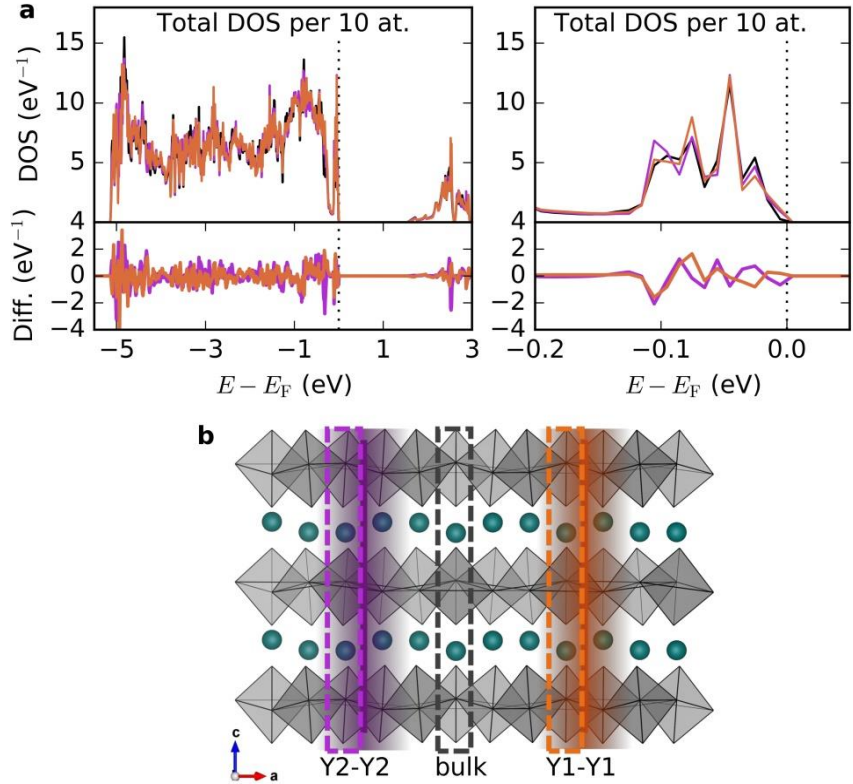

**Figure S3 | Electronic structure in domain and domain-wall regions. a**, The upper left panel shows the electronic density of states for three 10-atom sections of a 4x1x1 supercell of YMnO$_3$, indicated by colored rectangles in panel **b**. The difference between the domain wall density of states (DOS) and the domain center DOS is plotted in the lower panels of **a**. Panels on the right show a zoom-in to emphasize the electronic structure at the Fermi level. **b**, The section marked by the black dashed line corresponds to the middle of a domain, while purple and orange dashed lines lie close to the Y2 and Y1 terminated neutral domain walls, respectively.

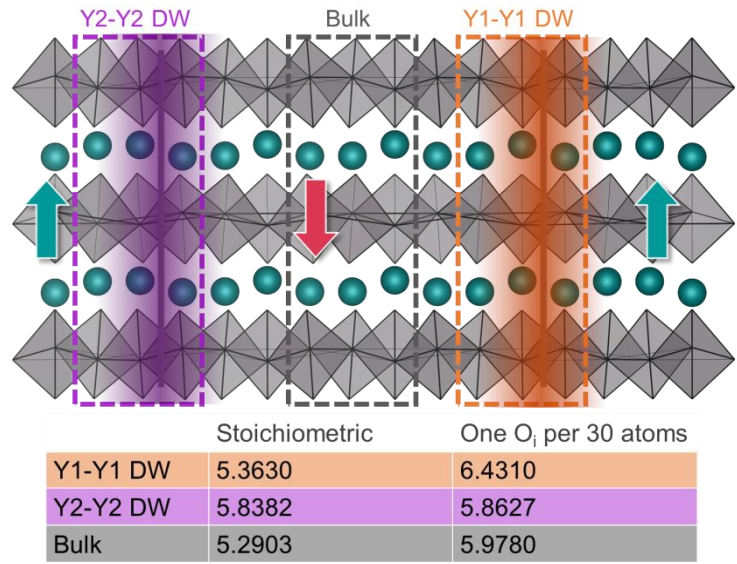

| | Stoichiometric | One O$_i$ per 30 atoms |
|---|---|---|
| Y1-Y1 DW | 5.3630 | 6.4310 |
| Y2-Y2 DW | 5.8382 | 5.8627 |
| Bulk | 5.2903 | 5.9780 |

**Figure S4 | Calculated dielectric constant for domain and domain-wall regions. a,** In order to estimate local variations in the dielectric constant, we first relax 5x2x1 supercells of YMnO$_3$ (YMnO$_3$ is isostructural and chemically similar to ErMnO$_3$, and the absence of f-electrons simplifies the density functional theory (DFT) calculations), with two neutral domain walls (Y1-Y1 and Y2-Y2). The arrows indicate the polarization directions, and the domain walls are indicated by thick colored vertical lines and shaded areas to show the extent of the structural distortions close to the walls. Relaxations of the stoichiometric supercell and of supercells containing one interstitial oxygen at a selection of stable positions, shown by orange spheres in Fig. 5c, were performed. Then we extract six different unit cells, representing the two neutral domain walls (Y1-Y1 domain wall and Y2-Y2 domain wall) and the bulk, as illustrated by the colored dashed boxes, with and without interstitial oxygen, and perform static calculations to obtain the $\varepsilon_{zz}$. **b,** Dielectric constant $\varepsilon_{zz}$ of YMnO$_3$ calculated for stoichiometric supercells and with oxygen interstitials. The calculations show no statistical relevant variation in $\varepsilon_{zz}$, which suggests that neutral domain walls and bulk exhibit similar dielectric constants in hexagonal manganites (note that our calculations address relative changes while underestimating the absolute value of the dielectric constant ($\varepsilon_{bulk} \approx 20$) as Landau-corrections are not taken into account).

## 2. Supplementary Tables

| Point defect at domain wall | ΔE = E$^f_{DW}$ - E$^f_{bulk}$ (eV) |
|---|---|
| Interstitial oxygen @ Y1 DW | -0.036 |
| Interstitial oxygen @ Y2 DW | -0.028 |
| Oxygen vacancy @ Y1 DW | 0.120 |
| Oxygen vacancy @ Y2 DW | 0.120 |

**Table T1 | Segregation energy of oxygen point defects to neutral domain walls.** Formation energies for oxygen vacancies are calculated from $E^f_{v_O} = E_{YMnO_{3-\delta}} - E_{YMnO_3} + \mu_O$ while oxygen interstitials according to $E^f_{O_i} = E_{YMnO_{3+\delta}} - E_{YMnO_3} - \mu_O$, with chemical potential of oxygen $\mu_O$ = -4.5 eV. Segregation energies are calculated as the difference between the formation energy at the domain wall and the energy for the same defect in the middle of the domain. Formation energies for oxygen interstitials were calculated using a 5x2x1 supercell (this work), while oxygen vacancy calculations were performed with a 8x1x1 supercell (primary results to be published elsewhere). Computational details concerning magnetic order, relaxation criteria, functional and k-point sampling are found in the manuscript.